\documentclass[letterpaper, 10 pt, conference]{IEEEtran}

\usepackage{amsthm}
\usepackage{amsmath,amssymb}
\usepackage{booktabs,multirow}
\usepackage{siunitx}
\usepackage{graphicx}

\usepackage{algorithm}
\usepackage[noend]{algpseudocode} 

\usepackage[inkscapelatex=false]{svg}

\usepackage{upgreek}
\usepackage{stmaryrd}

\usepackage{color}

\usepackage[pdfa,hidelinks]{hyperref}
\usepackage{optidef}
\usepackage{graphics} 
\usepackage{graphicx}
\graphicspath{{figures/}}
\usepackage{subcaption}
\usepackage{amsmath} 
\usepackage{amssymb}  
\usepackage{balance}
 
\usepackage{verbatim}

\usepackage{bm}

\DeclareMathOperator*{\argmin}{arg\,min}

\makeatletter
\newcommand\gobblepars{%
    \@ifnextchar\par%
 {\expandafter\gobblepars\@gobble}%
{}}
\makeatother

\def\wham#1{\medbreak\pagebreak[3]%
\noindent\textbf{#1}\ \ \gobblepars}

\newcounter{rmnum}

\newlength{\noteWidth}
\long\def\notes#1{\ifinner
             {\tiny #1}
             \else
             \marginpar{\parbox[t]{\noteWidth}{\raggedright\tiny #1}}
             \fi}
             
           \def\notes#1{\typeout{See notes!}}
\def\archive#1{}

\def\MC{\text{\MC}}

\def\transpose{{\hbox{\it\tiny T}}}

\def\argmin{\mathop{\rm arg\, min}}

\newcommand{\field}[1]{\mathbb{#1}}

\def\Re{\field{R}}

\def\eqdef{\mathbin{:=}}

\newtheorem{theorem}{Theorem}[section]

\newtheorem{proposition}[theorem]{Proposition}
\newtheorem{lemma}[theorem]{Lemma}

\usepackage{cleveref}

\crefname{algorithm}{Algorithm}{Algorithms}
\Crefname{algorithm}{Algorithm}{Algorithms}

\Crefname{corollary}{Corollary}{Corollaries}
\Crefname{eqnarray}{eq.}{eqs.}
\Crefname{equation}{eq.}{eqs.}

\Crefname{figure}{Fig.}{Figs.}
\Crefname{tabular}{Tab.}{Tabs.}
\Crefname{table}{Tab.}{Tabs.}
\Crefname{lemma}{Lemma}{Lemmas}

\Crefname{theorem}{Thm.}{Thms.}
\Crefname{definition}{Definition}{Definitions}
\Crefname{section}{Section}{Sections}
\Crefname{proposition}{Prop.}{Propositions}
\Crefname{assumption}{Assumption}{Assumptions}
\Crefname{example}{Example}{Examples}

\def\barell{{\overline {\ell}}}

\def\bfmath#1{{\mathchoice{\mbox{\bfmath$#1$}}%
{\mbox{\boldmath$#1$}}%
{\mbox{\boldmath$\scriptstyle#1$}}%
{\mbox{\boldmath$\scriptscriptstyle#1$}}}}

\def\Tol{\text{\rm Tol}}

\def\cX{c_{\text{\tiny X}}}
\def\cU{c_{\text{\tiny U}}}

\def\cdG{c_{{\text{\lower1pt\hbox{r}}}} }

 \def\FRAC#1#2#3{\genfrac{}{}{}{#1}{#2}{#3}}

\def\ddt{{\mathchoice{\FRAC{1}{d}{dt}}%
{\FRAC{1}{d}{dt}}%
{\FRAC{3}{d}{dt}}%
{\FRAC{3}{d}{dt}}}}

\def\ddtp{{\mathchoice{\FRAC{1}{\partial}{\partial t}}%
		{\FRAC{1}{\partial}{\partial t}}%
		{\FRAC{3}{\partial}{\partial t}}%
		{\FRAC{3}{\partial}{\partial t}}}}

\def\ddr{{\mathchoice{\FRAC{1}{d}{dr}}%
{\FRAC{1}{d}{dr}}%
{\FRAC{3}{d}{dr}}%
{\FRAC{3}{d}{dr}}}}

\def\half{{\mathchoice{\FRAC{1}{1}{2}}%
{\FRAC{1}{1}{2}}%
{\FRAC{3}{1}{2}}%
{\FRAC{3}{1}{2}}}}

\def\clB{{\cal B}}
\def\clC{{\cal C}}

\def\clH{{\cal H}}

\def\clT{{T}}
\def\clU{{\cal U}}

\def\cX{c_{\text{\tiny X}}}

\def\cdG{c_{{\text{\lower1pt\hbox{d}}}} }

\usepackage{bbold} 

\RenewEnviron{BaseMiniExclam}[7]{%
	\selectConstraintMult{#1}%
	\renewcommand{\localOptimalVariable}{#2}%
	\begin{subequations}
		\ifthenelse{\equal{#7}{b}}{\allowdisplaybreaks}%
		#4
		\begin{alignat}{5}
		\bodyobj{#2}{#3}{#6}{#5}
		\BODY
		\end{alignat}
	\end{subequations}%
	\setStandardMini
}

\def\fee{\upphi}

\makeatletter
\newcommand{\oset}[3][0ex]{%
  \mathrel{\mathop{#3}\limits^{
    \vbox to#1{\kern-2\ex@
    \hbox{$\scriptstyle#2$}\vss}}}}
\makeatother

\def\xcev{\oset[-.1ex]{\shortleftarrow}{x}} 
\def\zetacev{\oset[-.1ex]{\shortleftarrow}{\zeta}}
\def\ucev{\oset[-.1ex]{\shortleftarrow}{u}}

\def\uH{\underline{H}}

\def\st{\text{\rm s.t.\,}}

\def\tHor{{\hbox{\tiny$\mathcal{T}$}}} 
\def\sHor{{\hbox{\footnotesize$\mathcal{T}$}}} 
\def\Hor{{\mathchoice{\mathcal{T}}{\sHor}{\tHor}{\tHor}}}

\def\clHcev{{\oset[-.2ex]{\shortleftarrow}{\mathcal{H}}}}
\def\clCcev{{\oset[-.2ex]{\shortleftarrow}{\mathcal{C}}}}

\def\uHcev{{\oset[-.2ex]{\shortleftarrow}{\underline{H}}}}
\def\psicev{{\oset[-.2ex]{\shortleftarrow}{\psi}}}
\def\Psicev{{\oset[-.2ex]{\shortleftarrow}{\Psi}}} 

 \def\imp{\xi}

 \def\MPCt{\tau}
 \def\MPCshift{t_s}
 \def\INITt{t_0}

\makeatletter
\def\thanks#1{\protected@xdef\@thanks{\@thanks
		\protect\footnotetext{#1}}}
\makeatother
\title{Forecast and Model Predictive Control of Distributed Energy Resource Aggregators for Net-Demand Balancing}
\author{Obai~Bahwal, Oliver~Kosut, and Lalitha~Sankar%
\thanks{O.~Bahwal, O.~Kosut, and L.~Sankar are with the School of Electrical, Computer and Energy Engi
neering, Arizona State University, Tempe, AZ 85281, USA (emails: obahwal@asu.edu, okosut@asu.edu, lsankar@asu.edu).}%
}

\begin{document}

\maketitle


\begin{abstract}
With the rapid demand for energy, even the incorporation of bulk renewable energy sources is not entirely sufficient to meet demand besides adding supply uncertainty.
Distributed Energy Resource Aggregators (DERAs) have the potential to address this uncertainty via aggregation and control of decentralized distributed energy sources, thereby acting like virtual power plants. We present a new approach that combines forecasting and model-predictive control to assign DERAs to follow net-demand patterns, while accounting for the dynamics of the aggregate energy sources and their capacity limits. Each DERA is represented as a flexible ``virtual battery" with constraints on state-of-charge and power limits. The dispatch problem is set up as a long-term model predictive control task that aims to minimize differences from desired charge levels, output ramping, and net-load tracking errors. To keep operations efficient in real time, we implement a rolling-horizon MPC, which updates decisions regularly using the latest marginal-demand forecasts. For forecasting, we present two models: linear regression and long-short term memory (LSTM) neural network. Using high-resolution CAISO net-demand data and five typical DERA types, our simulations demonstrate how well our approach tracks marginal-demand; in particular, we highlight the tradeoffs between forecasting horizon times and MPC update rate as well as the dependence on the choice of the load forecasting model. Our results also indicate a slight edge for LSTM models over linear regression for desired time shifts and horizon choices. 

\end{abstract}

\section{Introduction}

\IEEEPARstart{M}{assive} renewable generation introduces volatility and uncertainty in the grid, leading to sharp ramps and disturbances in net-demand (demand minus renewable generation). Traditionally, these ramps and disturbances have been addressed using fossil fuel-based generators such as coal and natural gas, which are costly, inefficient, and not environment friendly.  
 
 The grid operator can tackle these challenges by intelligently allocating distributed energy resources (DERs), such as solar photovoltaics, wind turbines complemented with battery storage, and flexible loads. \emph{We propose a methodology that leverages aggregate DER models to optimally allocate DERs to meet the ramps and disturbances in marginal-demand forecasts (marginal-demand is net-demand minus bulk generation) using a model predictive control (MPC) framework}.  To this end, we model the dynamical behavior of each aggregated DER type as a battery with capacity constraints. The goal of the DER aggregator (DERA) is then to learn the optimal control policy over time to meet the demands of the grid. 


The grid operator and DERAs form the agents in our proposed approach. Aggregators are entities that engage with the consumers to aggregate DERs  (e.g., EnergyHub, Enel X, and utility companies). 
The role of the grid operator (e.g., ERCOT and CAISO) is to coordinate DERAs and bulk generation to procure and dispatch the resources required to balance supply and demand. The formulation in this paper focuses on resource allocation across different DERAs. However, the precise way in which the control is designed at the individual device level to achieve tracking is beyond the scope of this paper (see \cite{matbusmey23} and the references therein for individual-level control).

The control architectures presently used in the grid focus primarily on short-term optimization \cite{dorfler2019distributed}. Traditionally, economic dispatch with the participation of large generation plants is performed as a static optimization problem \cite{woowolshe13}. However, the optimal dispatch of DERAs needs to account for the dynamics of these aggregations, thereby requiring an optimal control framework. However, for sufficient DERA control, we need to plan for the long term despite uncertainties in net-demand, thus, reliable long-term forecasting that accounts for short-term dynamics is essential.
Moreover, even as the accuracy of forecast information becomes less reliable over long timescales, the challenge of jointly using both (more reliable) short-term and (less reliable) long-term forecasts to optimize real-time DER allocations while maintaining grid-level optimality appears to be unsolved.


\wham{Our Contributions} We address the above mentioned challenges by proposing a model predictive control (MPC) framework that captures the dynamics of the DERAs. For the MPC to design optimal control, it in turn requires real-time load forecasting. To this end, we first present multiple marginal-demand forecast models for different horizon times. These models rely on previous day marginal-demand and calendar indicators for prediction. To evaluate the importance of load forecasting, we consider two distinctly different models: (i) a simple linear regression (LR) model, and (ii) a more complex long short-term memory (LSTM) neural network (NN).  

We leverage our forecast model to introduce our key contribution: an MPC framework with constraints on the state of charge (SoC),  aggregate capacity of each DER type, and a net power balance constraint across all agents. The objective of the MPC is to minimize the net cost of DER generation under the above mentioned constraints for a forecast time horizon and repeatedly over time shifts. The resulting optimal policy allows for short-term optimization while taking into account relatively longer-term forecast. MPC approximates the solution to a long-run optimal control problem by solving a sequence of tractable, finite-horizon, and therefore finite dimensional optimization problems. 

We present extensive numerical results on both the forecast and MPC aspects. We first compare the performance of the two types of forecast models (LR and LSTM) we consider across different prediction horizon times. For both LR and LSTM-based forecasts, we compare the results of the MPC to identify the best predictor across both horizon lengths and time-shifts. Our simulations clearly highlight the importance of: (i) well-trained forecasting models, (ii) MPC horizon times, and (iii) frequency (time-shift) of MPC updates. \textit{Our extensive results show that this framework allows very close tracking of loads to minimize mismatch while actively meeting ramping, power, and state of charge constraints.  } 


\wham{Related Research}\label{sec: related research}
MPC has been used in specific grid contexts; in fact, it has strong guarantees for transient operations and ensures constraint satisfaction, while also accounting for long-run optimality \cite{rawlings2017model}. In \cite{ref:scalable} and \cite{ref:network}, MPC is used at the distribution-level, in \cite{ref:scalable} with a focus on market profit, and in \cite{ref:network} with a focus on security and resilience. 
On the other hand, we focus on transmission-level marginal-demand balancing based on forecast and MPC integration. DER aggregation is considered in \cite{ref:gridaware,ref:Hierarchical} for distribution-level control, using convex inner approximation, and ancillary service reserves, while we rely on marginal-demand forecast aligned to MPC for optimal DER allocation at the grid-operator level.
While all these approaches share in common with ours the goal of minimizing a chosen objective for real-time resource allocation, our work is distinct by including (a) dynamic models for DER aggregations; (b) hard constraints on SoC, power balance, and battery capacity constraints; (c) integrating multi-horizon marginal-demand (load) forecasting; and (d) empirical evaluation over multiple time-shifts for the MPC optimization.  We note that our approach does not focus on the downstream allocation aspects considered in \cite{ref:scalable,ref:network,ref:gridaware,ref:Hierarchical}.

Finite-horizon convex optimization formulations for the control of DERAs are discussed in \cite{espalm20}: a centralized quadratic program is solved to generate different command signals for the different classes of DERs. In \cite{bencolmal19}, the authors solve a non-linear AC optimal power flow (OPF) with the participation of an aggregation of DERs, specifically thermostatically controlled loads; the problem is modeled as a Markov decision process (MDP) and is reformulated as a finite-horizon convex program. Similar formulations are 
considered in \cite{cammatkiebusmey18, matmeybalans21}. 
Related work also studies when battery-like aggregation preserves feasibility and optimality for heterogeneous DER schedules \cite{ref:optimality}, and how uncertainty-aware aggregate feasible regions and equivalent operating costs can be constructed for large populations of DERs \cite{ref:aggop}. However, these finite-horizon formulations and aggregation models do not explicitly analyze forecast-coupled control loop in a multi-horizon and multi-time shift (update frequency) settings.
Unlike traditional finite-horizon approaches that overlook real-time disruptions and the balance between forecast length and update frequency, we integrate forecast models across several timeframes with a rolling-horizon MPC under constraints, and analyze how the forecast horizon time $\MPCt$ and update interval $\MPCshift$ affect tracking performance and outcomes. 

We note that our work in \cite{matangkossan24} (which has some overlap with this paper) formally introduces the MPC setup used here; however, it uses the true loads for learning the allocation policies. Such data is never available non-causally, and thus, one needs a reliable load forecasting model. One can therefore view this prior work as a simple blueprint. The novelty of this work relative to \cite{matangkossan24} is in extensive experimental results using CAISO load data to (i) design two distinctly different (in complexity and capabilities) load forecasting models and (ii) integrate their predictions into the MPC framework to learn the real-time resource allocation policy for the DERAs. 

\wham{Notation}  
\\
\noindent
$\MPCt$: control time horizon for an MPC iteration indexed by $t \in \{\INITt,t_0+1,\ldots,\INITt+\MPCt-1\}$.\\
$\INITt$: starting time for an MPC iteration.\\
$\MPCshift$: time shift for $t_0$ between successive MPC iterations.\\
$\ell(t)$ : forecast of marginal-demand at time $t$. 
\\
$M$: number of DERAs, indexed by $i \in \{1,\ldots,M\}$.
\\
$x_i (t)$ : SoC for DERA $i$ (units: GWh). \\
$p_i(t)$ : power output of aggregator $i$ (units: GW).\\
$u_i(t)$: is the ramping input, i.e., $u_i(t) \eqdef p_i(t+1) - p_i(t)$\\
$x(t)$: vector with the $i$-th entry $x_i(t)$. Similar notation for $p(t)$ and $u(t)$.\\ 
$\epsilon(t)$ : tracking mismatch error of marginal-demand.\\
$\kappa^\text{\tiny X}_i$: design parameter for DER ramping costs.\\
$\kappa^\text{\tiny U}_i$: design parameter for DER state costs.\\
$\star$ indicates an optimal solution, e.g. $x_i^\star(t)$ is the optimal state of the $i$-th aggregation at time $t$.\\
Parenthesis are used for time indices, subscripts are used to enumerate resource aggregations, and superscripts are used to denote iterations of the optimization algorithm: e.g., $x^k_i(t)$ refers to the $k$-th iteration of the $i$-th aggregation's SoC at time $t$.\\
For ease of notation, time indices are dropped for signals to denote the entire trajectory when appropriate; e.g., $p_i$ denotes the power trajectory of the $i$-th aggregation, i.e., $\{p_i(t): t \in \{\INITt,\ldots,\INITt+\MPCt-1\}\}$. Throughout the paper, we use the terms demand and load interchangeably.  


\section{Problem Setup}
\label{s:prelim}
We begin by describing the DERA model via aggregate DER dynamics, the forecasting methodology, control objective, and set up the MPC optimization framework. We note that our dynamical model for DERAs as well as the MPC objective functions largely mimics the setup in \cite{matangkossan24}. 

\subsection{Generalized linear battery models}
\label{s:ves_models}

We adopt linear aggregate models for DER dynamics, an approach that is well established in prior work (see \cite{haosanpoovin15, hugdompoo16, ma2011decentralized,matangkossan24}). 
Each DERA is represented by a generalized battery model with known power and state-of-charge (SoC) limits. For the $i$-th aggregation, $i \in \{1,\ldots,M\}$, the discrete-time dynamics and constraints are
\begin{subequations}
\label{e:DER_model}
\begin{align}
\label{e:SoC_ODE}
x_i(t+1)  & = \alpha_i x_i(t)   - \beta p_i(t),\\ 
\label{e:SoC_cap}
|x_i(t)| & \le C_i, \\
\label{e:Power_cap}
-\eta_i^- & \le p_i(t) \le \eta_i^+, 
\end{align}
\end{subequations}
with the following definitions for the $i$-th aggregation:
\begin{itemize}
    \item $x_i(t)\in \Re$ is the SoC, The nominal
    SoC is defined to be 0 (nominal refers the behavior of the DERA when the DERs are not participating in dispatch or regulation).
    \item $\beta$ denotes the sampling time constant.
    \item $\alpha_i\in [0,1]$ is a leakage parameter. For thermostatically controlled loads (e.g., air conditioners, water heaters), $\alpha_i$ relates to the thermal time constant, which equals $\beta/(1-\alpha_i)$.
    \item $p_i(t) \in \Re$ is the power delivered at time $t$ by the \hbox{$i$-th} DERA, expressed as a deviation from a baseline. Hence $p_i(t)$ may be positive even for purely consuming devices (e.g., an air conditioner “supplies’’ power when it consumes less than baseline).
    \item $C_i$ is the SoC capacity limit.
    \item $\eta_i^+$ and $-\eta_i^-$ are the positive and negative power limits, respectively, where $\eta_i^+,\eta_i^- > 0$.
\end{itemize}

For a brief discussion on how the power and energy capacity limits are obtained for aggregations of grid-responsive loads, we refer the reader to our prior work in \cite{matangkossan24}. 


\subsection{Modeling Costs of Deviations}
\label{s:ves_models}

For DERA $i$, the penalty for deviating from nominal state is denoted $c_i(x_i):\Re \to \Re_+$, and is chosen to be strongly convex and twice differentiable function. The resulting aggregate cost of state deviations is
\begin{equation}
\cX(x) \eqdef  \sum_{i=1}^{M}   c_i(x_i).
\label{e:cX}
\end{equation}
While our formulation above is general, for ease of analysis and tractability of the optimization, we choose $c_i(\cdot)$ as the quadratic function 
\begin{equation}
    c_i(x_i) = \half \kappa^\text{\tiny X}_i x_i^2.
\end{equation}

We next define the ramping variable for DERA $i$ as
\begin{equation}
    \label{e:DER_ramping}
u_i(t) \eqdef p_i(t+1) - p_i(t),
\end{equation}
and penalize rapid changes to discourage policies that lead to spiky power use. 
To this end, we introduce the total ramping cost as: 
\begin{equation}
\cU(u) \eqdef  \sum_{i=1}^M \half \kappa^\text{\tiny U}_i u_i^2,
\label{e:cU}
\end{equation}
where $\kappa^\text{\tiny U}_i \ge 0$ are parameters that represents the cost of ramping DERAs up or down.

Later in the paper, we introduce specific classes of DERs and present values for these constants based on prior empirical studies for such DERs.

Resource allocation must satisfy a power-balance constraint, i.e., the total power from all aggregations should match the (baseline-normalized) marginal-demand up to a controllable mismatch. We introduce a slack variable $\epsilon(t)$ to capture this mismatch as:
\begin{equation}
\epsilon(t)  = \ell(t) - \barell - \sum_i p_i(t),
    \label{e:supply=demand}
\end{equation}
where $\ell(t)$ is the marginal-demand forecast at time $t$ and $\barell$ is a known baseline. We capture the cost of this mismatch via the penalty $c_\epsilon(\epsilon)$, which we assume is strongly convex and twice differentiable.
For the rest of the paper, we consider a quadratic penalty on the mismatch $\epsilon$, i.e.,
\begin{equation}
\label{eq:mismatch}
    c_\epsilon(\epsilon(t)) = \half \kappa_\epsilon \epsilon(t)^2 
\end{equation}
where $\kappa_\epsilon$ is a cost parameter.


\subsection{Model Predictive Control}
\label{s:mpc}

In theory, one can pose the allocation problem as an \emph{infinite-horizon} optimal control problem with running costs comprising the state, ramping, and mismatch penalties given by:
\begin{equation}
\lim_{\clT\to\infty} \sum_{t=0}^{\clT-1}    \left[    \cX(x(t)) + \cU(u(t)) + c_\epsilon(\epsilon(t)) \right].
    \label{e:obj}
\end{equation}
The initial states $x_i(0)$ are assumed known. The problem is subject to the dynamics \eqref{e:DER_model}–\eqref{e:DER_ramping} and the balance constraint \eqref{e:supply=demand}.

While the optimization in \eqref{e:obj} can lead to elegant theoretical solutions, it is not feasible in practice due to: 
(i) limited reliability of long-horizon time models/forecasts, and (ii) the presence of capacity constraints. 
MPC addresses these challenges by optimizing over a short, look-ahead moving window while enforcing the same physical and operational limits \cite{rawlings2017model}. To this end, we refine the finite-horizon MPC optimization framework, introduced in \cite{matangkossan24} for this setting\footnote{Since we focused on a finite horizon setting, we removed the terminal penalty cost which was computed in \cite{matangkossan24} using the solution to the Riccati equation assuming infinite horizon optimization.} to obtain the following optimization detailed below.

Let $\MPCt$ denote the prediction horizon time and $\MPCshift\!\le\!\MPCt$ the time shift between successive optimizations. For an iteration starting at time $\INITt\!\ge 0$ with state $x(\INITt)\in\Re^M$, the finite-horizon problem is
\begin{mini!}
	{x, p, u, \epsilon}{ \begin{aligned} \begin{split}
	\sum_{t=\INITt}^{\INITt+\MPCt-1} \!\!\big[\cX(x(t)) + \cU(u(t)) + c_\epsilon(\epsilon(t)) \big]
 \end{split} \end{aligned} \label{qp19obh}}
	{\label{qp19}}{}
	\addConstraint{\ell(t) - \barell - \sum_i p_i (t) }{=\epsilon(t) \label{e:balancecons}}
	\addConstraint{ {x}_i(t+1)}{= \alpha_i x_i(t) - \beta p_i(t) \label{e:soccons}}
 \addConstraint{ p_i(t+1)}{= p_i(t) + u_i(t) \label{e:powercons}}
	\addConstraint{-C_i}{\leq x_i(t)\leq C_i \label{e:SoCcap}}
	\addConstraint{-\eta^-_i}{\leq p_i(t)\leq \eta^+_i \label{e:PowerCap}}{}
\end{mini!}

\section{Design Implementation}
\label{s:dd}

This section outlines the algorithmic approach used to realize the objective in \eqref{qp19obh} subject to the aggregate DERA dynamics \eqref{e:DER_model}–\eqref{e:DER_ramping} and the balance constraint \eqref{e:supply=demand}. The design integrates a moving-window marginal-demand forecast with a  Model Predictive Control (MPC) formulation that repeatedly solves a finite-horizon convex program using the most recent forecast of marginal-demand with horizon time matching the MPC.

\subsection{Forecast Implementation}
\label{ss:fcst}


To provide accurate marginal-demand forecasts aligned with MPC, we consider two types of learning models for forecasting. Specifically, we consider either a simple linear regression (LR) model or a long short-term memory (LSTM) model. For such each type (LR or LSTM), we construct multiple forecast models, each with a distinct look-ahead (horizon time) that match the MPC’s look-ahead window. That is, the forecast models predict the load at all times (in $\beta$ increments) between the present and the end of the MPC horizon time. 

Every forecast uses a feature set built from the previous day’s marginal-demand up to the current time $t=\INITt$, the day of the week, the time of day, and a national-holiday binary indicator. At each MPC update time $t=\INITt$, we assemble a feature vector $\phi(\INITt)$ from these values. The forecasting model then predicts the marginal-demand for the future MPC-selected horizon time, denoted~$\MPCt$.


\wham{LSTM:} Similar to LSTM implementations in \cite{ref:lstm1}, \cite{ref:lstm2}, \cite{ref:lstm3}, and \cite{ref:lstm4}, our model utilizes a single channel input layer with the features described above, an encoder layer consisting of LSTM layer and fully-connected layer with rectified linear unit (ReLU), and a decoder layer that outputs the time-series prediction for $\MPCt$ time steps into the future in one shot.

\wham{LR:} The input to our LR model is the same set of feature vectors as input to the LSTM. However, since LR can only predict a single point in the horizon time $\MPCt$, we train a bank of LR models for each point from the first future sampling point until the end of the horizon time $\MPCt$.

\begin{algorithm}[t]
\caption{Rolling Window MPC Operation}
\label{alg:rhm_pc}
\begin{algorithmic}[1]
\Require $\beta$; $\MPCt$; $\MPCshift\!\le\!\MPCt$; $\barell$; costs $\cX$, $\cU$, $c_\epsilon$; $x(\INITt)\in\Re^M$
\While{MPC ON}
  \State \textbf{accumulate} previous loads at times $\{t-24h, ...,\INITt\}$.
  \State \textbf{Forecast} the horizon time $\MPCt$
  \State \textbf{Optimize} by solving \eqref{qp19}  
  \State \textbf{Apply} the first $\MPCshift$ trajectories
  \State \textbf{Advance} $\INITt \gets \INITt + \MPCshift$ 
\EndWhile
\end{algorithmic}
\end{algorithm}

\begin{table}[b]
\vspace{-0.75 em}
	\centering
	\label{tab:RA}
	\begin{tabular}{|| l c c c c c r ||}
		Par. & Unit & DER1 & DER2 & DER3 & DER4 & DER5 \\
		Type & --- & ACs & E-WHs & bldgs & RFGs & EVs \\
		N & million & 10 & 10 & 1 & 10 & 1   \\
		$\alpha_i$ & ---  & 0.98 & 0.99 & 0.97 & 0.96 & 0.99  \\
		$C_i$ & GWh &  8 & 5 & 2.3 & 5 & 50  \\
            $\eta^+_i$ & GW & 20 & 4 & 103 & 2 & 3.6 \\
            $\eta^-_i$ & GW & 30 & 50 & 3 & 3 & 3.6\\
            $\kappa^\text{\tiny X}_i$ & --- & 1 & 2 & 5 & 5 & 2 \\
            $\kappa^\text{\tiny U}_i$ & --- & 1e9 & 1e9 & 1e9 & 1e9 & 1e9 \\
            $\kappa_\epsilon$ & --- & 1e8 & 1e8 & 1e8 & 1e8 & 1e8 \\
            $\beta$ & s & 300 & 300 & 300 & 300 & 300
	\end{tabular}
	\bigskip
\caption{Parameters for each class of DERA, corresponding to the discrete-time linear dynamical model in \eqref{e:DER_model}.}
\end{table} 

\subsection{Integration of Forecast and MPC Algorithm}
\label{ss:forecast_mpc}
Algorithm~\ref{alg:rhm_pc} shows the implementation of the rolling window MPC incorporating the load forecast.
At each MPC start time $\INITt$, a short-term forecast of marginal-demand is obtained over the horizon time $\MPCt$ based on the previous 24 hour marginal-demand data. 
Then using this forecast, the algorithm solves a finite-horizon problem \eqref{qp19} over $[\INITt,\INITt+\MPCt]$. 
In this way, the forecast shapes the desired aggregate response. After solving \eqref{qp19}, only the first $\MPCshift$ control actions are applied. After time $\MPCshift$, the horizon is shifted forward by $\MPCshift$ step, new measurements are incorporated, and the forecast is updated before solving the optimization problem again. 

\begin{table*}[t]
\centering
\caption{Mean Squared Error Averaged over a Year}
\label{table:results}
\setlength{\tabcolsep}{2.5pt}
\scriptsize
\newcommand{\NA}{\multicolumn{1}{c}{---}}

\resizebox{\textwidth}{!}{%
\begin{tabular}{ll *{24}{S[table-format=1.9]}}
\toprule
& & \multicolumn{8}{c}{Actual vs DERA} \\
\cmidrule(lr){3-10}
Model & $\MPCshift$ & $\MPCt$
& {30m} & {1h} & {2h} & {3h} & {6h} & {12h} & {24h} \\
\midrule


LR   & 15m &
& 0.6170 & 0.5671 & 0.5228 & 0.4895 & 0.4235 & 0.4082 & 0.4076 \\

LR   & 30m &
& \NA & 0.9601 & 0.8769 & 0.7936 & 0.5992 & 0.5616 & 0.5605 \\

LR   & 60m &
& \NA & \NA & 1.1183 & 1.0285 & 0.8049 & 0.7552 & 0.7542 \\

LR   & 90m &
& \NA & \NA & 1.2538 & 1.1651 & 0.9398 & 0.8824 & 0.8814 \\

LR   & 120m &
& \NA & \NA & \NA & 1.2775 & 1.0478 & 0.9818 & 0.9810 \\

LR   & 150m &
& \NA & \NA & \NA & 1.3594 & 1.1298 & 1.0524 & 1.0515 \\

LR   & 180m &
& \NA & \NA & \NA & \NA & 1.1889 & 1.0944 & 1.0934 \\

\midrule


LSTM & 15m &
& 0.6521 & 0.6040 & 0.5988 & 0.5825 & 0.5359 & 0.5341 & 0.5314 \\

LSTM & 30m &
& \NA & 0.9180 & 0.8622 & 0.8056 & 0.6312 & 0.5946 & 0.6021 \\

LSTM & 60m &
& \NA & \NA & 1.0701 & 1.0050 & 0.8076 & 0.7527 & 0.7691 \\

LSTM & 90m &
& \NA & \NA & 1.1850 & 1.1195 & 0.9408 & 0.8727 & 0.9074 \\

LSTM & 120m &
& \NA & \NA & \NA & 1.1851 & 1.0175 & 0.9450 & 0.9802 \\

LSTM & 150m &
& \NA & \NA & \NA & 1.2565 & 1.0690 & 0.9974 & 1.0494 \\

LSTM & 180m &
& \NA & \NA & \NA & \NA & 1.1520 & 1.0391 & 1.1152 \\

\bottomrule
\end{tabular}%
}
\end{table*}

\section{Simulations}
\label{s:numerics}

\begin{figure*}[!t]
\centering
\includegraphics[angle=0, trim={0 80 0 80}, clip, width=\linewidth]{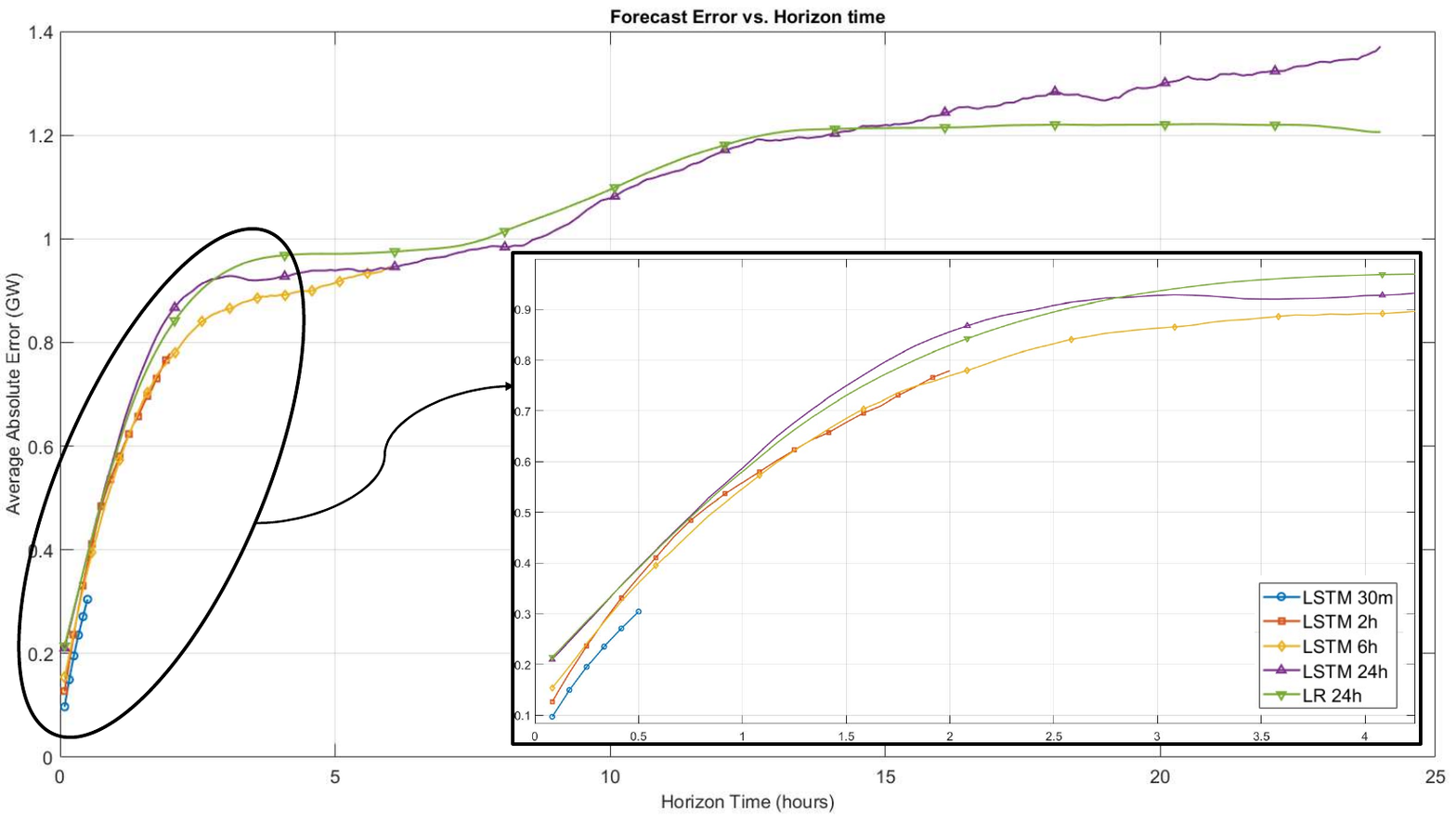}
\caption{\small This is a plot of the Mean Absolute Error (MAE) (y-axis) for every choice of horizon time step (x-axis) between the prediction (of either LR or LSTM) model relative to the true load. This average is computed for the period between August $2^{\text{nd}}$, 2024 until August $3^{\text{rd}}$, 2025 with $\MPCshift = 5\text{m}$ and $\MPCt = \{30\text{m}, 2\text{h}\, 6\text{h}, 24\text{h}\}$. For the LR model, only the $24\text{h}$ trace is shown as it captures the predictions of every (point-wise) linear model whereas for the LSTM, we learn different models for different prediction horizon times.}
\label{fig:fcsterr}
\end{figure*}

The simulations detailed here include five classes of DERs, each corresponding to an aggregation of millions of DERs. The DER classes are as follows: air conditioners (ACs), electric water heaters (E-WHs), building HVACs (bldgs), refrigerators (RFGs), and electric vehicles (EVs) with 100 kWh lithium ion batteries. Table~I provides details about the type and number of DERs as well as the model parameters for each aggregation. The number of DERs in each aggregation are extrapolated based on the population of California. We compute $\alpha$ values and the capacity limits for the different DERAs using data from the following papers: for ACs, E-WHs, and RFGs \cite{matdyscal15}, for bldgs \cite{hugdompoo16}, and for EVs \cite{ma2011decentralized}. 

We use real historical net-demand data from CAISO from August 2021 until August 2024 for training of forecast models. Moreover, we verify the robustness of our forecast models by running all of our experiments for a year on the period between August $2^{\text{nd}}$, 2024 and August $3^{\text{rd}}$, 2025 to ensure that the data contains all seasonal changes and (ir)regular load patterns over weekdays, weekends, and holidays. The CAISO data set obtained has a sampling resolution $\beta$ of 5 minutes, where 39 sample points are missing from the testing period. For those missing points, we interpolate the values by calculating the average spread between two known endpoints to conduct our simulations.

To distinguish the true demand from the forecast demand, we introduce the notation $\ell_a(t)$ for the actual marginal-demand. (As in the MPC formulation, $\ell(t)$ represents the forecast demand.) This is defined is the difference between actual net-demand and the baseline $\barell$. The baseline is obtained as a moving average over a period of one day based on the actual net-demand.

\begin{figure*}[t!]
    \centering
    \begin{subfigure}[t!]{0.5\textwidth}
        \centering
        \includegraphics[angle=0, trim={20 170 20 170 },clip, width=\linewidth]{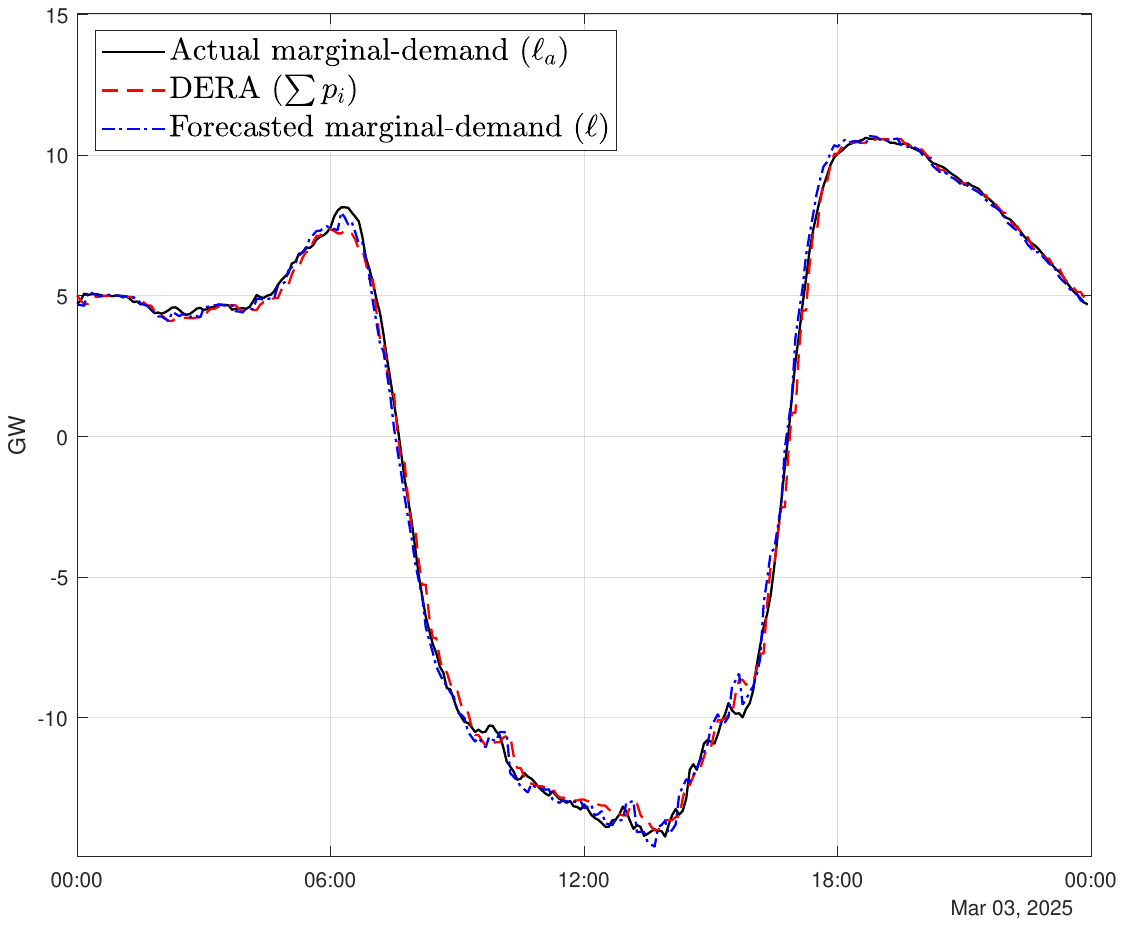}
        \caption{\small LR test with $\MPCt=24\text{h}$ and $\MPCshift=15\text{m}$.}
    \label{fig:lrbest}
    \end{subfigure}%
    ~    
    \begin{subfigure}[t!]{0.5\textwidth}
        \centering
        \includegraphics[angle=0, trim={20 170 20 170 },clip, width=\linewidth]{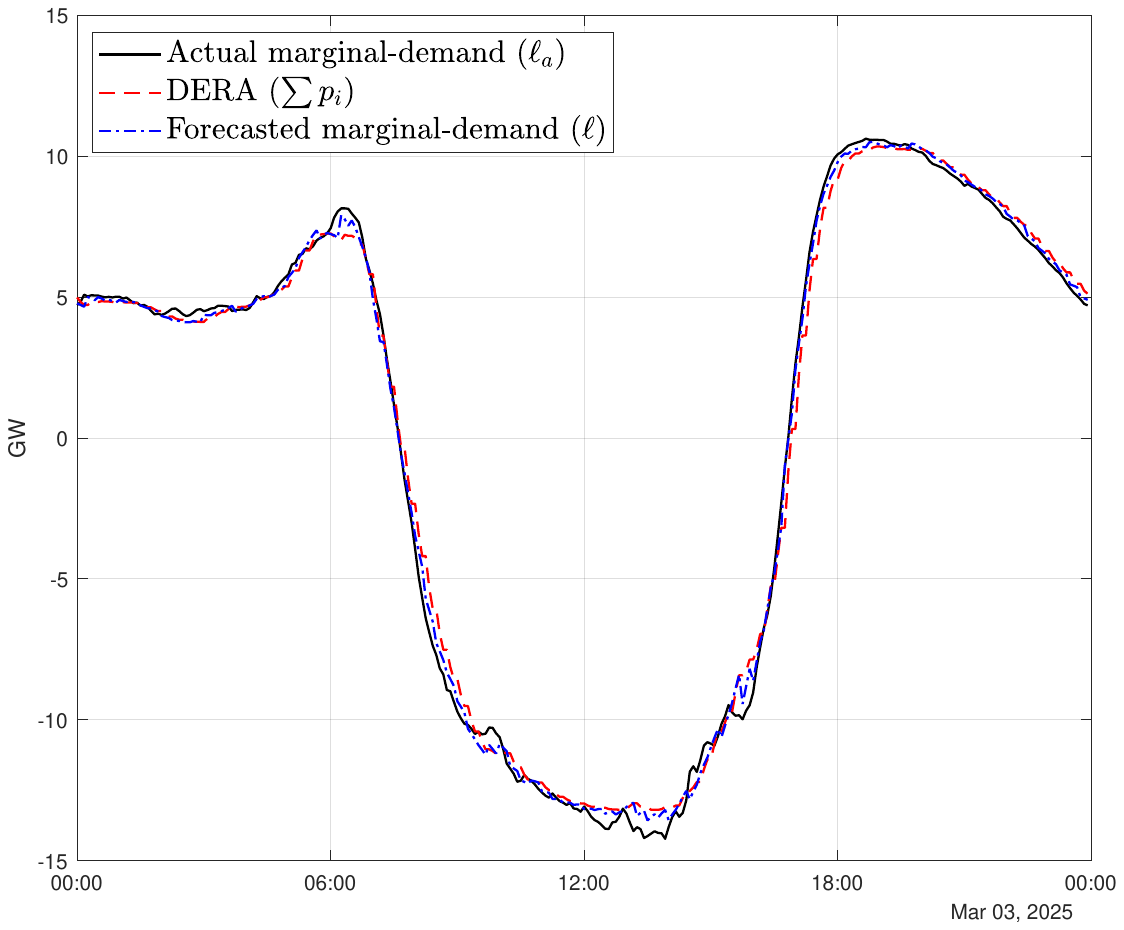}
        \caption{\small LSTM test with $\MPCt=24\text{h}$ and $\MPCshift=15\text{m}$.}
    \label{fig:nnbest}
    \end{subfigure}%
    
    \begin{subfigure}[t!]{0.5\textwidth}
        \centering
        \includegraphics[angle=0, trim={20 170 20 170 },clip, width=\linewidth]{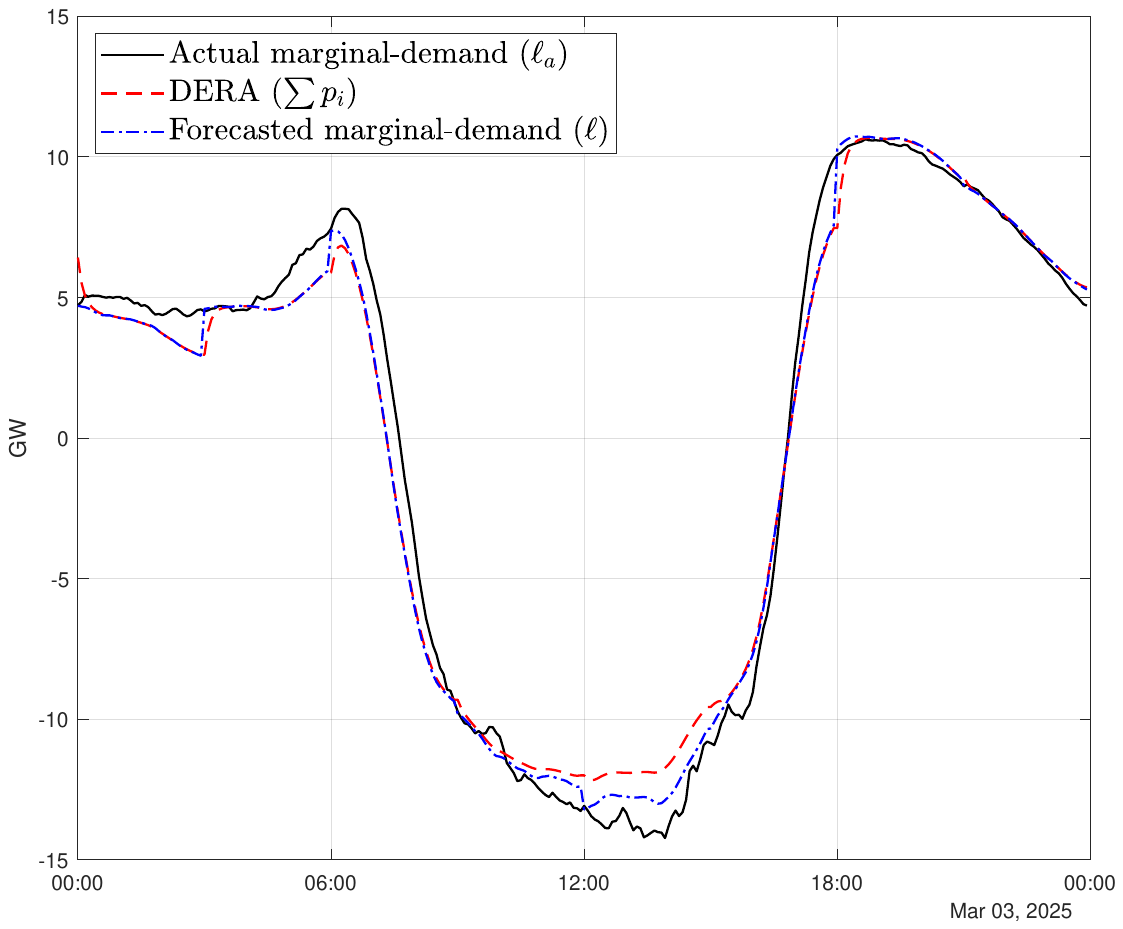}
        \caption{\small LR test with $\MPCt=24\text{h}$ and $\MPCshift=180\text{m}$.}
    \label{fig:lrworst}
    \end{subfigure}%
    ~ 
    \begin{subfigure}[t!]{0.5\textwidth}
        \centering
        \includegraphics[angle=0, trim={20 170 20 170 },clip, width=\linewidth]{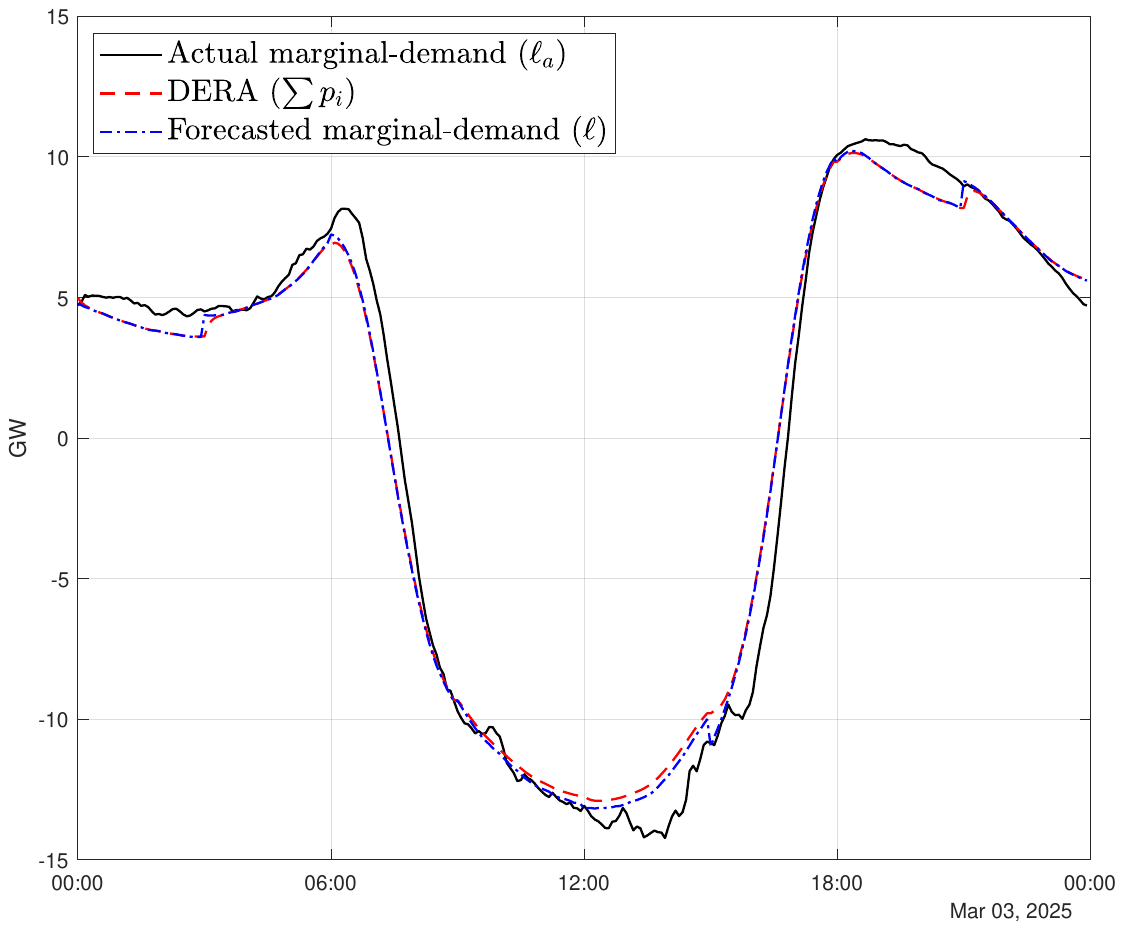}
        \caption{\small LSTM test with $\MPCt=24\text{h}$ and $\MPCshift=180\text{m}$. }
    \label{fig:nnworst}
    \end{subfigure}
    \caption{DERA tracking comparison of best performing $\MPCshift$ settings showing actual marginal-demand, DERA Allocation, and forecasted marginal-demand against worse performing settings for the simulation period between March $2^{\text{nd}}$, 2025 until March $3^{\text{rd}}$, 2025.}
\label{fig:signals}
\end{figure*}

\subsection{Forecast Models Validation}
\label{sec:forecast}
To evaluate the accuracy of our forecast models, we run our algorithm with a rolling window to forecast for the testing period. For each predictive model, we compute the absolute difference between forecast and actual marginal-demand for each point in $\{\INITt, \INITt+\MPCshift, ..., \MPCt\}$, and for each $\MPCt \in \{30\text{m}, 2\text{h}, 6\text{h}, 24\text{h}\}$. The forecast points are updated every $\MPCshift = 5\text{m}$ and the Mean Absolute Error (MAE) for each point in the horizon times is calculated. Figure \ref{fig:fcsterr} show the mismatch plots for $\MPCshift = 5\text{m}$ and $\MPCt = \{30\text{m}, 2\text{h}, 6\text{h}, 24\text{h}\}$.

We observe from Fig. \ref{fig:fcsterr} that the MAE plots that LSTM and LR exhibit similar behavior in that with increasing horizon time $\MPCt$, the error increases. This is expected as predicting the near future where loads are more correlated temporally should be more accurate. We note that that it suffices to consider a single LR trace for $24\text{h}$ as it only forecasts one future time point at a time; in other words, the LR model can be viewed as a bank of LR forecasters that predict a single point on the horizon time using the same features.

Our results in Fig.  \ref{fig:fcsterr} suggest that the LSTM models demonstrate consistently better performance for short horizon times such as those shown in the figure, i.e., 30 mins, as well as for 2 and 6 hours. For a horizon time of 24 hours, the models appear to be more comparable. We conjecture that an LSTM with more layers would perform better than LR for this 24-hour time-shift. We chose to train all of our LSTM models using the same NN structure for consistency and comparability, even though long-term forecasting, generally, require different model design compared to short-term forecasting.

\begin{figure*}[t!]
    \centering
    \begin{subfigure}[t!]{1\textwidth}
        \centering
        \includegraphics[angle=0, trim={70 230 70 230 },clip, width=\linewidth]{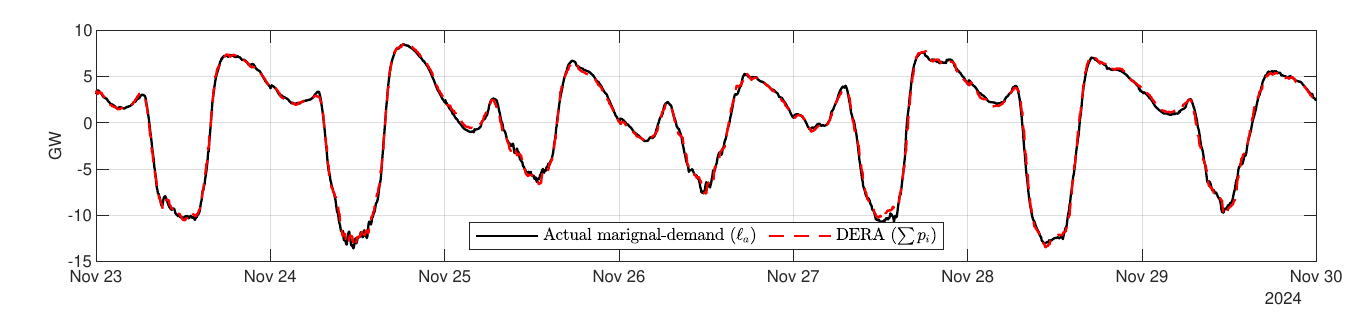}
        \caption{\small Cumulative DERA trajectory versus actual marginal-demand}
    \label{fig:compower}
    \end{subfigure}%
      
    \begin{subfigure}[t!]{1\textwidth}
        \centering
        \includegraphics[angle=0, trim={70 230 70 230 },clip, width=\linewidth]{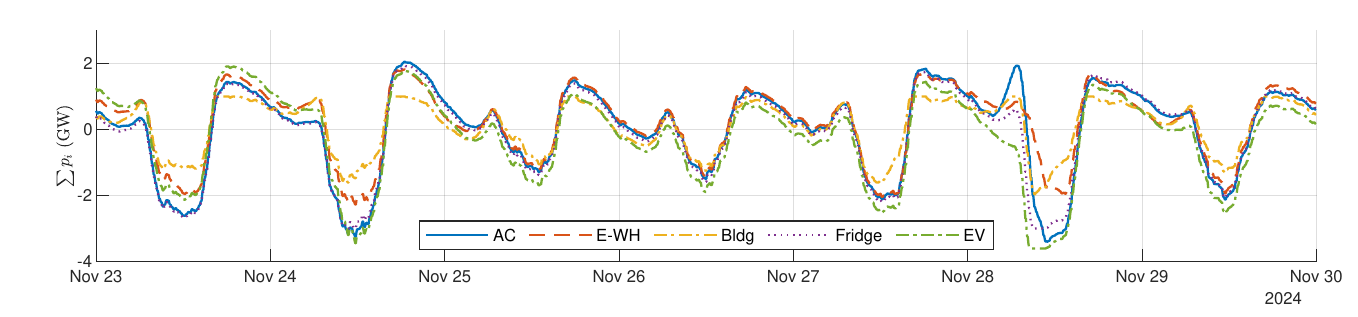}
        \caption{\small Individual DER power Trajectories}
    \label{fig:indvpower}
    \end{subfigure}%
    
    \begin{subfigure}[t!]{1\textwidth}
        \centering
        \includegraphics[angle=0, trim={70 230 70 230 },clip, width=\linewidth]{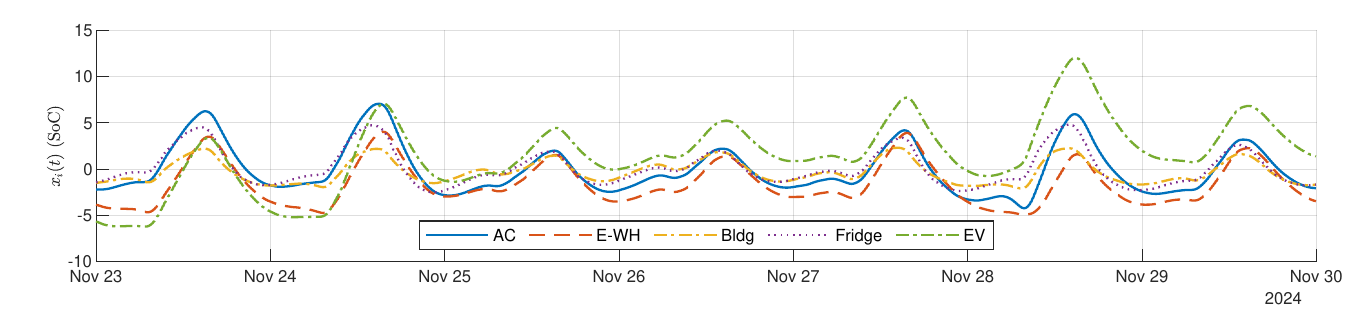}
        \caption{\small Individual SoC Trajectories}
    \label{fig:soc}
    \end{subfigure}%
    \caption{Illustration of regulation-type services capabilities of DERAs for the period between $23^{\text{rd}}$ of Nov. 2024 and $29^{\text{th}}$ of Nov. 2024 using LR forecasts with $\MPCt=24\text{h}$ and $\MPCshift = 15\text{m}$.}
\label{fig:regulations}
\end{figure*}

\subsection{MPC Results}

In Figs. \ref{fig:lrbest} and \ref{fig:nnbest}, we plot the following over a 24 hour period for Mar. 3, 2025 with $\MPCt=24\text{h}$ and $\MPCshift=15\text{m}$: (a) actual marginal-demand, (b) DERAs Allocation, (c) forecasted marginal-demand. We note that this choice of $\MPCt=24\text{h}$ and $\MPCshift=15\text{m}$, leads to strong performance by both LR and LSTM models. 
In contrast, Figures \ref{fig:lrworst} and \ref{fig:nnworst} show the same comparison (for the same day) for $\MPCt=24\text{h}$ and $\MPCshift=180\text{m}$ which clearly displays the degraded performance with longer $\MPCshift$.

In Figs \ref{fig:compower}, \ref{fig:indvpower}, and \ref{fig:soc} we show the cumulative power supplied by the DERs along with actual marginal-demand, the individual DER power trajectories, and the SoC trajectories, respectively. Figure \ref{fig:regulations} is plotted the week of Thanksgiving 2024 between $23^{\text{rd}}$ of Nov. 2024 and $29^{\text{th}}$ of Nov. 2024 to span over weekdays, weekends, and holidays. The results are obtained using LR model with $\MPCt=24\text{h}$ and $\MPCshift = 15\text{m}$. We aim to demonstrate with these plots the ability to provide regulation-type services utilizing DERAs in the presence of fluctuating power trajectories where we see the DERs charge during times of low marginal-demand (i.e., they consume more energy with respect to nominal), so that they can supply energy (consume less than nominal) when the marginal-demand ramps up.

In Table \ref{table:results}, we evaluate the efficacy of the DER aggregations that result from the optimal MPC policy using either an LR or LSTM forecast model. The metric we use is the mean squared error (MSE) between the DERAs allocation and the actual marginal-demand. Since the time shift $\tau$ cannot be larger than the horizon time, the table captures values for different time-shifts (rows of table) with increasing horizon times (columns) starting from that time shift. 

\begin{figure*}[t!]
    \centering
    
    \begin{subfigure}[t]{0.5\textwidth}
        \centering
        \includegraphics[angle=0, trim={120 100 120 100}, clip, width=\linewidth]{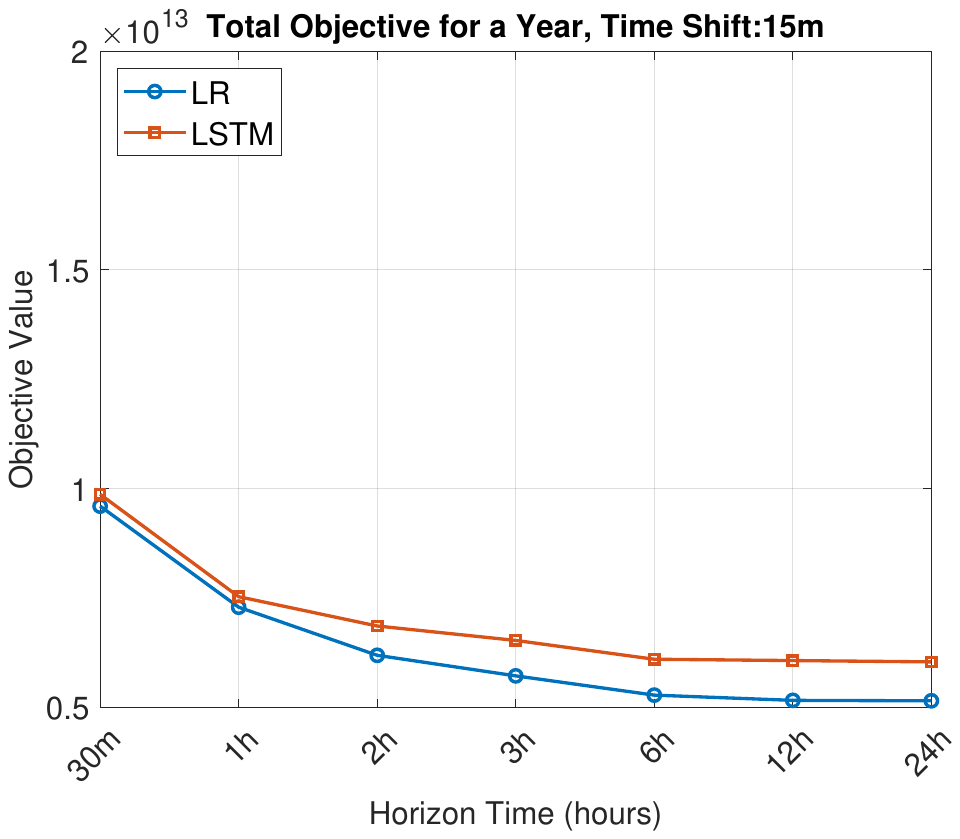}
        \caption{Comparison of objective with $\MPCshift=~15\text{m}$}
    \label{fig:obj15}
    \end{subfigure}%
    ~ 
    \begin{subfigure}[t]{0.5\textwidth}
        \centering
        \includegraphics[angle=0, trim={120 100 120 100}, clip, width=\linewidth]{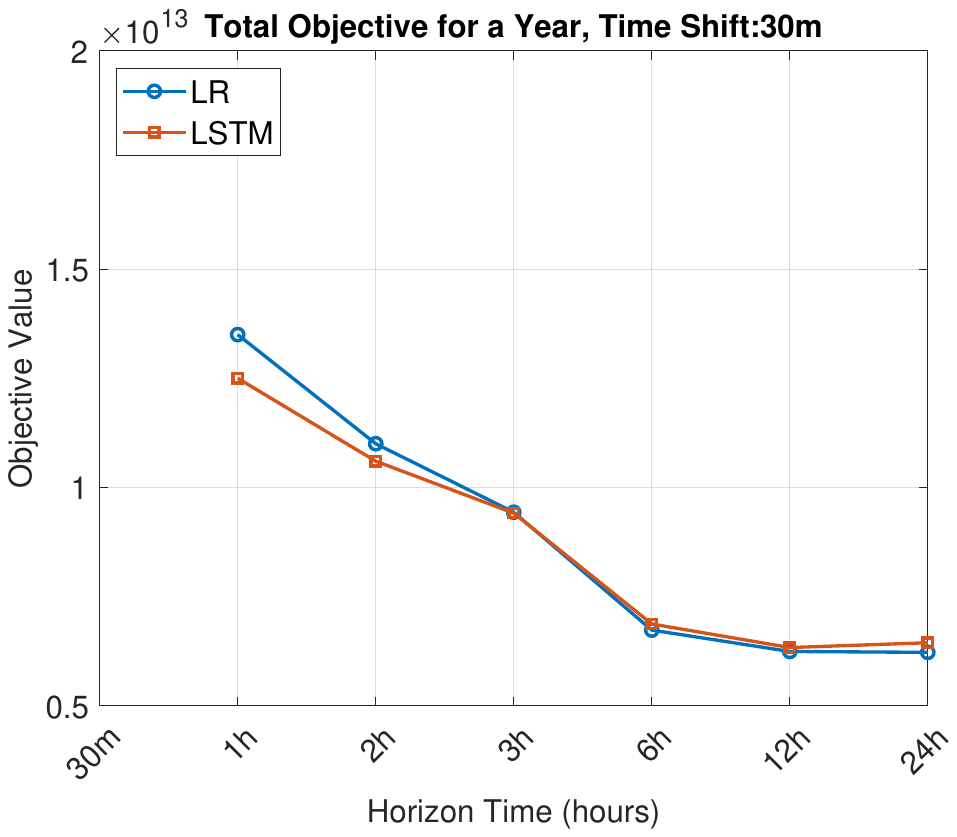}
        \caption{Comparison of objective with $\MPCshift=~30\text{m}$}
    \label{fig:obj30}
    \end{subfigure}
    \begin{subfigure}[t]{0.5\textwidth}
        \centering
        \includegraphics[angle=0, trim={120 100 120 100}, clip, width=\linewidth]{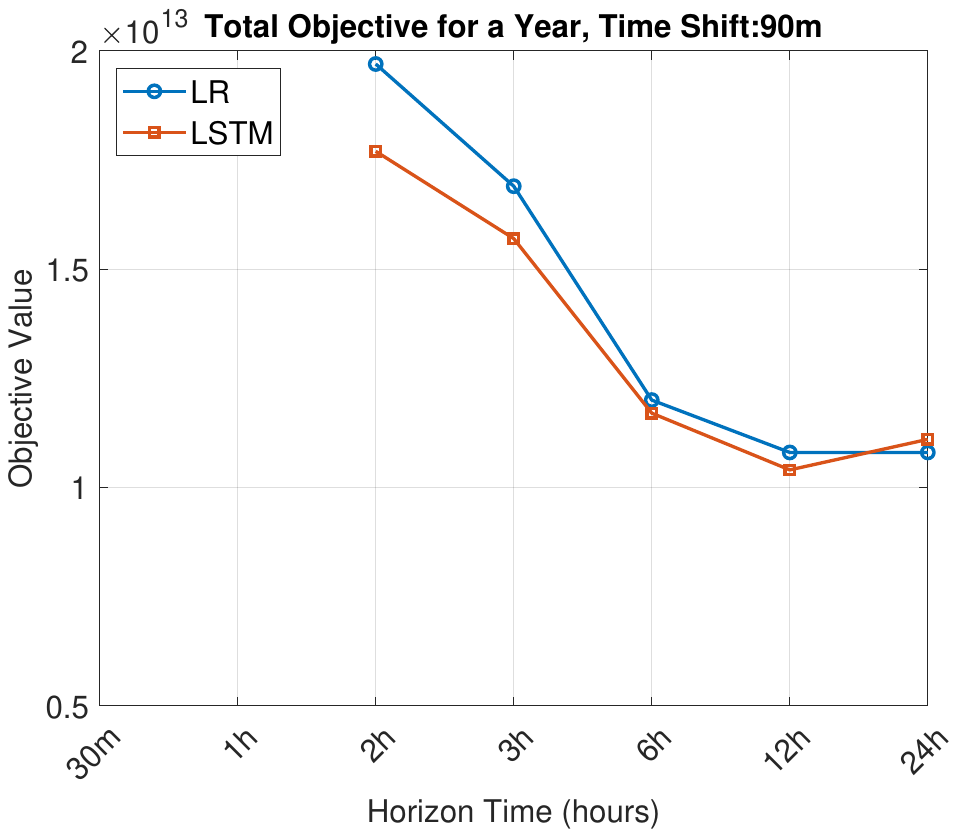}
        \caption{Comparison of objective with $\MPCshift=~90\text{m}$}
    \label{fig:obj90}
    \end{subfigure}%
    ~ 
    \begin{subfigure}[t]{0.5\textwidth}
        \centering
        \includegraphics[angle=0, trim={120 100 120 100}, clip, width=\linewidth]{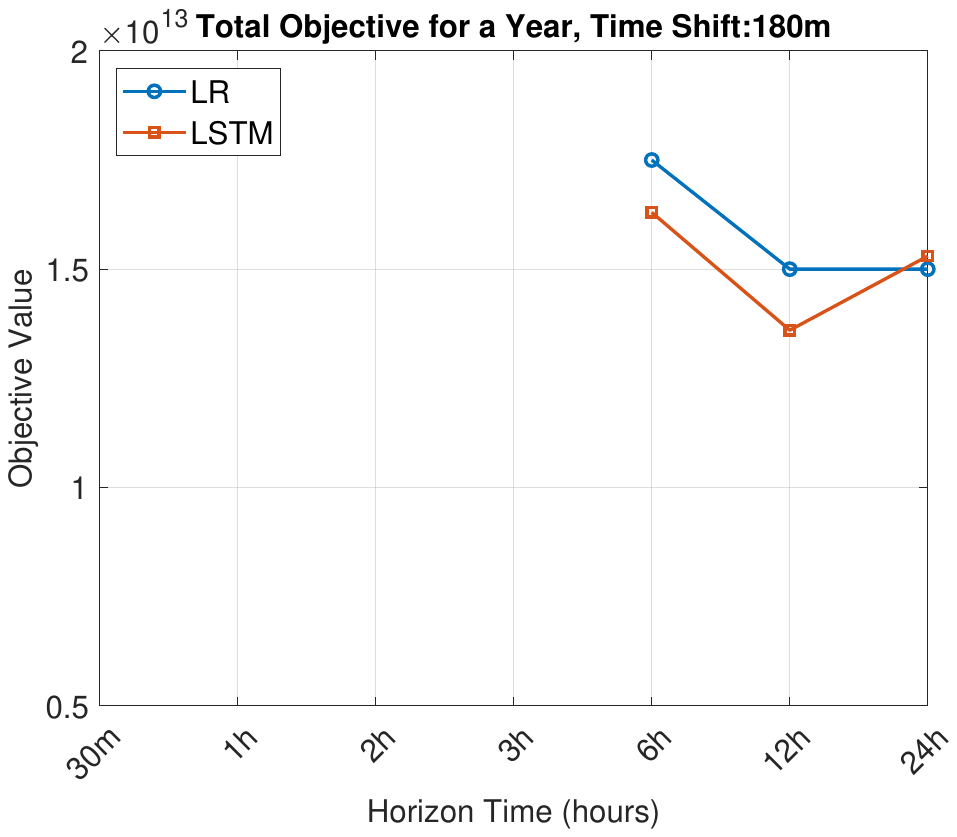}
        \caption{Comparison of objective with $\MPCshift=~180\text{m}$}
    \label{fig:obj180}
    \end{subfigure}
    \caption{Objective over the period between August $2^{\text{nd}}$, 2024 and August $3^{\text{rd}}$, 2025, for LR and LSTM compared to having the mismatch cost, $c_\epsilon$, calculated with reference to actual marginal-demand}
\label{fig:obj}
\end{figure*}

Each plot in Figure~\ref{fig:obj} shows the total value of the objective function in \eqref{qp19obh}, when computed using the true loads, i.e., instead of marginal-demand forecast $\ell(t)$ in \eqref{e:supply=demand}, we use the actual marginal-demand $\ell_a(t)$. This objective is calculated by summing all objective costs: $\cX$, $\cU$, $c_\epsilon$. Thus, the total objective captures not only how well the DERA output matches the load, but it also accounts for state-of-charge and ramping costs. These plots evaluate the objective function with the actual marginal-demand, rather than the forecast as is used in the MPC itself, in order to evaluate the quality of the DERA control once it has been dispatched in the real system.

In particular, Figs. \ref{fig:obj15}, \ref{fig:obj30}, \ref{fig:obj90}, and \ref{fig:obj180} show the optimization objective in \eqref{qp19obh} over the entire 365-day testing period, i.e., $T=365$, between August $2^{\text{nd}}$, 2024 and August $3^{\text{rd}}$, 2025, for time-shifts of $\MPCshift=15, 30, 90, 180$ mins, respectively.  



\subsection{Observations}
We collate our results into two key observations as detailed below.

(i) As seen in both Table~\ref{table:results} and Figure~\ref{fig:obj}, the overall performance of the proposed MPC has better performance with decreased time-shift $\MPCshift$ and increased horizon time $\MPCt$. We see these effects in Table~\ref{table:results} and Figure \ref{fig:obj} as a lower MSE between actual marginal-demand and aggregator response and a  lower objective value for smaller time-shifts, respectively. 
Both these effects are expected for MPC, which will work better when executed more often (shorter time-shifts), or with longer-term forecasts (longer horizon times) to plan with. These observations hold even though, in the case of the LSTM forecast model, the forecast of a particular time in the future is slightly worse when the horizon time is longer, as seen in Fig.~\ref{fig:fcsterr}.


(ii) The overall performance of the MPC is better with the LR forecast than that with LSTM forecast for shorter $\MPCshift$, while LSTM performs better at longer $\MPCshift$. This pattern is breached for $\MPCt=24\text{h}$ and that is likely due to the increased error for the LSTM forecaster, as seen in subsection \ref{sec:forecast}, at the longest horizon time, which in turns affects the MPC planning for DERA allocation. The general phenomenon that the MPC with LR performs better at short time-shifts --- even though the LR forecast is less accurate than the LSTM over short time intervals --- seems to result from the fact that with a short time-shift, the MPC has a greater ability to continuously update its plan, and so the accuracy of the forecast models, where their accuracy are relatively similar, does not matter as much as, perhaps, other characteristics of the forecast. These characteristics may include smoothness or the sign of the mismatch between forecast and actual marginal-demand. In particular, compared to the LSTM forecast, the direction of the error of the LR forecast more frequently matches the direction of the mismatch in the MPC.

\section{Conclusions and Future Work}
We have shown the efficacy of MPC-based control design for the optimal allocation of DERAs to meet marginal-demand. Our MPC approach has two distinct advantages: (i) it uses more accurate marginal-demand forecasts available in the short-term for resource allocation, and (ii) it ensures constraint satisfaction, while still accounting for long-run optimality. Complex forecasting models like LSTM are especially effective for predicting over longer time-shifts, as shown by their strong performance under those settings. In contrast, simpler LR models are better suited to shorter time-shifts. For system operators, depending on the application, a tradeoff between LSTM and LR involves balancing the need for long-term planning against the benefits of short-term accuracy. LSTM is preferable when longer time-shifts (less frequent updates) forecasts are needed, while LR offers a lightweight solution, when shorter time-shifts (more frequent updates) are needed, that excels at closely tracking short-term marginal-demand changes without demanding significant computational resources. Future problems include integrating a more comprehensive MPC problem by adding network limitations, market objectives, individual device level control of DERAs, and other grid considerations.

\bibliographystyle{IEEEtran}
\bibliography{strings,markov,q,extras,PolicyCollapseExtras,CollapseExtras}  

\end{document}